\documentclass[12pt]{article}
\usepackage{amsmath,latexsym,amssymb,epsfig,psfrag}
\newcommand{\be}{\begin{equation}}
\newcommand{\ee}{\end{equation}}
\newcommand{\I}{\mathrm i} 
\newcommand{\ba}{\begin{array}} 
\newcommand{\ea}{\end{array}}
\newcommand{\bal}{\begin{align}}
\newcommand{\eal}{\end{align}}
\newcommand{\bqa}{\begin{eqnarray}}
\newcommand{\eqa}{\end{eqnarray}}
\newcommand{\nms}{\normalsize}

\newcommand{\abs}[1]{\left|#1\right|}

\newcommand{\tam}[1]{{\text {\Large $#1$}}}

\newcommand{\lb}{\left(}
\newcommand{\rb}{\right)}
\newcommand{\lsb}{\left[}
\newcommand{\rsb}{\right]}
\newcommand{\lcb}{\left\{}
\newcommand{\rcb}{\right\}}

\def\pd{\partial}
\def\spd{\slash\hspace{-.2cm}\partial}
\def\di{{\rm d}}
\def\a{\alpha}
\def\b{\beta}

\def\g{\gamma}
\def\gd{\g^{\dot 5}}

\def\d{\delta}

\def\m{\mu}
\def\n{\nu}

\def\th{\theta}

\def\th{\theta}
\def\l{\lambda}
\def\L{\Lambda}

\def\w{\omega}

\def\sig{\sigma}
\def\Sig{\Sigma}
\def\e{\epsilon}

\def\id{{\bf 1}}


\begin{document}
\begin{titlepage}  

\vskip 2cm
\begin{center}  
\vspace{0.5cm} \Large {\sc ON THE EQUIVALENCE BETWEEN
\\REAL AND SUPERFIELD 5D FORMALISMS}

\vspace*{1.5cm}
  
\normalsize  
  
{\bf 
D.~Diego~\footnote[1]{david.diego@nd.edu}$^{\,a,b}$}

\smallskip   
\medskip   
{$a$~\it Department of Physics, 225 Nieuwland Science Hall, U. of Notre Dame,}\\ 
{\it Notre Dame, IN 46556-5670, USA.}

\vspace{.2cm}

{$b$~\it Theoretical Physics Group, IFAE}\\ 
{\it E-08193 Bellaterra (Barcelona), Spain}\\



\vskip0.6in \end{center}  
   
\centerline{\large\bf Abstract}  

\vspace{.5cm}

We explicitly prove the equivalence and construct a dictionary between two different supersymmetric formalisms for five-dimensional theories commonly used in the literature. 
One is the real formalism, which consists in doubling the number of degrees of freedom and then imposing 
reality constraints and the other is the usual superfield formalism. 
\noindent

   
\vspace*{2mm}   
  
\end{titlepage}  
%
%
%
%
%
\section{\Large Introduction}
%
%
The superspace description of 4D $N = 1$ supersymmetry was firstly introduced by Wess and Zumino in 1974 and quickly developed by other authors~\cite{Ferrara:1974ac}. 
It consists of a representation of the supersymmetric algebra on the so called superspace: a manifold spanned by the 
usual spacetime coordinates plus a set of anticommuting spinorial variables ($\theta, \bar\theta$). The irreducible representations of supersymmetry are now functions 
on the superspace called superfields, which contain the dynamical degrees of freedom plus auxiliary fields furnishing a representation of the superalgebra, and invariant actions can be easily constructed projecting the 
(product of) superfields onto $\theta^2\bar\theta^2$ component or $\th^2$ ($\bar\th^2$) if they are chiral (anti-chiral) superfields~\cite{wb}.
 For extended supersymmetries, $N \geq 2$, the irreducible representations can be split into $N = 1$ superfields and invariant actions can be constructed the same way by requiring them to respect the global $R$-symmetry of the algebra.
 For instance, in 5D $N=1$ supersymmetry, which corresponds to $N = 2$ from the 4D point of view, the matter hypermultiplet can be decomposed into two 4D-chiral superfields\footnote{This is not always the case. For $N=1$ in 6D there is no off-shell formulation for the
 (massive) matter hypermultiplet. The reason is that the minimal (massive) multiplet with $1/2$ as maximum helicity in 6D should be equivalent to an 5D $N=2$ multiplet, and hence, charged under a central generator of the algebra. 
 However, there is no central charge for $N=1$ in 6D~\cite{Lee:2003mc}. For massless representations, however, the central charge realizes trivially on the physical states.} while the vector hypermultiplet can be expressed as one 4D-chiral 
superfield and one 4D-vector superfield~\cite{Mirabelli:1997aj}. \newline
As we already pointed out, with the superfield formulation one can build up invariant actions in a rather systematic way, which is a great advantage in writing invariant couplings with respect to the component field formulation. In this sense, the aim of the
 present paper is to 
shed some light on the formalism used in Refs.~\cite{de Wit:1984px,Diego:2005mu}. More precisely, the starting point  is a model defined 
in a 5D manifold with 4D boundaries~\cite{Diego:2005mu} (the so called interval approach) and mass like terms strictly localized on the latter\footnote{Being thus equivalent to an orbifold scenario with bulk odd masses.}
and where all the fields are subject to reality constraints, the dictionary shall consist then in rewriting the whole action in terms of superfields~\cite{Barbieri:2002ic}.\newline
 This translation, as we will see, helps us to better understand 
the process of supersymmetry breaking by boundary terms. \newline
The paper is organized as follows:  In section 2 we will quickly review the model. In section 3 we will present the real-to-superfield dictionary, that is: the invariant action will be rewritten in terms of superfields 
and section 4 will be devoted to the supersymmetry breaking pattern. Finally, in section 5 we present our conclusions and an appendix is added at the end of the paper just containing some more technical details of the translation.
%
%
%
%
%
%
%
%
%
\section{\Large Real Formalism Revisited}\label{real}
%
    As it was mentioned in the introduction, the model is defined in a flat 5D manifold with boundaries: $\Sig = M_4\times I$, $M_4$ 
    being the 4D Minkowski space, $I$ the interval $[0, \pi R]$ and $R$ the compactification radius.  
    The metric signature is taken as $\eta_{MN} = {\rm diagonal}\lb +1,-1,-1,-1,-1\rb$ and
    the field content of the hypermultiplet in 5D is $(\Phi_i,\Psi,F_i)$ where $\Phi_i$ are complex scalars and $F_i$ auxiliary fields, both transforming as doublets of $SU(2)_R$ while  $\Psi$ is a Dirac fermion. 
 To have a manifest  $SU(2)_R$ covariance in the superalgebra we use the $N = 2$ 5D structure~\cite{Zucker:2003qv}
      \be
        \left\{Q_i\,,\,Q_j\right\} = \e_{ij}\,\g^M C\,P_M + \e_{ij}Z\,C\,,\label{n2algebra}
      \ee
     subject to a symplectic Majorana (SyM) constraint
     \be
     \bar Q^i \equiv Q_i^\dagger\g^0 =  \e^{ij} Q_j^T C\,,\label{sym}
     \ee  
     where $\e^{ij}$ is the total antisymmetric tensor and 
      \be
   C = -\id\otimes \I\sig_2 = \lb\ba{cc}-\I\sig_2&0\\ 0&-\I\sig_2\ea\rb\,,
   \ee
 is the 5D charge conjugation matrix verifying $C\,\g^M\,C = -\left(\g^M\right)^T$. In addition $\g^M = \lb\g^\m, \g^{\dot 5}\rb$ with\footnote{The dotted index stands for a Lorentz one and hence $\g^{\dot 5} = - \g_{\dot 5}$} $\g^{\dot 5} = - \I\g^5$ and
 {\small$\g^5 = \lb\ba{cc}\id&0\\0&-\id\ea\rb$}.
     Finally, $P_M$ are the spacetime translation generators, $Z$ is a central charge and  
     consistency with (\ref{sym}) imposes $Z$ to be hermitian.
     Now it is clear how the real formalism is implemented: we double the number of degrees of freedom and impose reality constraints, that is
\be
    \mathbb
H^\alpha=(\Phi_i,\Psi,F_i)^\a ,
\label{hp}
\ee
where $i$ is an $SU(2)_R$ index while $\a$ is an extra $SU(2)_H$ index. Those constraints read
\be
\bar\Psi_\alpha \equiv \left(\Psi^\a\right)^\dagger\g^0 =
\epsilon_{\alpha\beta}(\Psi^\beta)^TC\,,
\label{dirac}
\ee
\be
\bar\Phi_\a^i\equiv (\Phi_i^\alpha)^*=
\epsilon^{ij}\epsilon_{\alpha\beta}\Phi^{\beta}_j\,,
\label{scalars}
\ee
which can be compactly written as
\begin{align}
  \bar\Psi = &- \Psi^T \e\otimes C \,,\nonumber\\
  \Phi^* =& \e\otimes\e\,\Phi \,.\nonumber
\end{align}
The auxiliary fields verifying the same constraint as the scalars. In (\ref{scalars}) $\e_{\a\b}$ is again the total antisymmetric tensor and in both cases ($H$ and $R$) the convention taken is such that $\e^{12} = \e_{12} = 1$. \newline The action is given by 
%
\begin{align}
\mathcal S &=\int_\Sig\left( -\frac{1}{2}\bar\Phi\;
\partial^2 \Phi + \frac{\I}{2} \bar \Psi\gamma^M\partial_M \Psi+2\bar F
F
+2\I\bar F \mathcal M \Phi + \frac{1}{2}\bar \Psi
\mathcal M \Psi\right)\nonumber\\
\nonumber\\
&+\int_{\partial \Sig} \left(\frac{1}{4}\bar \Psi S_f\Psi
+\frac{1}{4}(\bar \Phi R_f \Phi)'+\frac{1}{4} \bar\Phi
N_f(-\id+R_f)\Phi\right)\,,
\label{actionbd}
\end{align}
where $\mathcal M, S_f, R_f \equiv T_f\otimes S_f$ are hermitian matrices and $N_f$ are real constants, $\mathcal M$ and $S_f$ act on $SU(2)_H$ indices while $T_f$ act on $SU(2)_R$. The subscript $f$ takes the values $0,\pi$ and indicates the boundary and 
the prime stands for the derivative with respect to the fifth coordinate. The reality constraints ensure the reality of the kinetic term since
\begin{align}
\bar\Psi\g^M\pd_M\Psi = - \Psi^T C\g^M\pd_M\e\Psi &= \pd_M\Psi^T\e\otimes C\g^M\Psi = - \pd_M\bar\Psi\g^M\Psi \,,\label{tder}\\
\bar\Phi\pd^2\Phi & = \pd^2\bar\Phi\,\Phi\,,
\end{align}
and for the rest of the terms to be real it is required that
$\mathcal M, S_f, T_f \in \emph{su}(2)$, thus, we can define (dimensionless) $3$-vectors $\vec p,\vec s_f, \vec t_f$ such that $\mathcal M  = M \vec p\cdot\vec\sig$ and $S_f = \vec s_f \cdot\vec\sig\,,T_f=\vec t_f\cdot\vec\sig$, where $M$ is a 
constant\footnote{$M$ can be suitably redefined such that $\vec p$ is a unit vector.} with dimension of 
energy and $\vec\sig$ are the Pauli matrices.
%
%
%
\subsection{\large Equations of motion and boundary conditions}
The variational principle applied to the bulk $+$ brane action yields a 5D variation (whose vanishing yields the equations of motion) plus a boundary one, the latter being
\be
  \frac{1}{2}\int_{\pd\Sig} \d\bar\Psi\lb-\I\g^{\dot 5} + S_f\rb\Psi+\d\bar\Phi'\lb\id + R_f\rb\Phi +\d\bar\Phi\lb-\id + R_f\rb\lsb\Phi'+ N_f \Phi\rsb\,, \nonumber
\ee
%
 and thus yielding the boundary conditions
 \bqa
    \left.\left(\id - \I \g^{\dot 5}\otimes S_f\right)\,\Psi\right|_f & = & 0\,, \label{fer}\\
    \left.\left(\id + R_f\right)\,\Phi\right|_f & = & 0\,,  \label{bos}\\
     \left(\id - R_f\right)\left[\Phi' + N_f \Phi\right]_f & = & 0\,.\label{bos'}
 \eqa
A necessary condition for those restrictions to allow a non trivial solution is that the matrices 
 $$\id + \I \g^{\dot 5}\otimes S_f$$
 \begin{center}
 {\rm and}
 \end{center} 
 $$\left(\ba{cc}\id + R_f &0\\ N_f\left(\id - R_f\right)&\id - R_f \ea\right)\,,$$
 must be singular\footnote{An exhaustive study of the spectrum allowed by these boundary conditions is made in Ref.~\cite{Diego:2005mu}.}. Their determinants are easily found to be
 $\left(1 - \abs{\vec s_f}^2\right)^2$ and $\left(1 - \abs{\vec s_f}\abs{\vec t_f}\right)^4$, respectively, and hence
$\abs{\vec s_f} = \abs{\vec t_f}=1$. On the other hand, the equations of motion for the auxiliary fields are
\be F = -\frac{\I}{2}\mathcal M\,\Phi\,,\qquad \bar F = \frac{\I}{2}\bar \Phi\,\mathcal M\,. \ee
%
%
\subsection{\large Supersymmetry of the action and boundary conditions}
The realization of the supersymmetric algebra (\ref{n2algebra}) at the level of the fields reads  
\begin{eqnarray}\label{susytransf}
    \delta_\chi\Phi^\alpha_i & = & \I\bar\chi_i\Psi^\alpha \,,\nonumber \\
    \delta_\chi\Psi_\alpha & = & - \gamma^M\chi^i\partial_M\Phi^\alpha_i + 2\chi^iF^\alpha_i \,,\nonumber \\
   \delta_\chi F^\alpha_i & = & -\frac{\I}{2}\bar\chi_i\gamma^M\partial_M\Psi^\alpha \,,
\end{eqnarray}
whose parameter satisfies an analogous symplectic Majorana reality constraint\footnote{Indeed it is needed to ensure the consistency of the supersymmetric algebra with the reality constraints of the fields.}
\be\bar\chi^i \equiv \lb\chi_i\rb^\dagger \g_0= \e^{ij} \chi_j^T C\,.\label{susyreal}\ee
As we show in Appendix~\ref{B} the boundary conditions are stable under supersymmetry if, and only if, {\nms $ \lb\I\gd - T_f^T\rb\chi=0$} and {\nms $N_f = M\,\vec p\cdot\vec s_f$}, where the projection on 
the supersymmetric parameters forces $\vec t_0 = \vec t_\pi$ in order to have a non vanishing residual supersymmetry.\newline
Concerning the supersymmetry of the action, the bulk and boundary pieces are not separately invariant, instead the bulk action varies into a total derivative which after partial integration combines with 
the boundary variation to give 
%
%
%
%
%
%
\begin{align}
\int_{\partial\Sig}&\lsb2\I\bar\Psi\g^{\dot 5}\chi\lb F + \frac{\I}{2}\mathcal M \Phi\rb + \frac{1}{2}\bar\Phi\lb\id + R\rb\d_\chi\Phi'+ \frac{1}{2}\lb\bar\Phi' +N\bar\Phi\rb\lb-\id + R\rb\d_\chi\Phi\right.\nonumber\\
         & \left. \frac{1}{2}\bar\Psi\lb-\I\g^{\dot 5} + S\rb\d_\chi\Psi\rsb\,,\label{susyvar}\end{align}
which cancels upon the use of boundary conditions, provided they are stable under supersymmetry, and the equations of motion for the auxiliary fields. This is to be expected since being the boundary term 
on-shell\footnote{There is no auxiliary field present although it is a mass term.}, the supersymmetry requires the boundary conditions to be satisfied~\cite{Belyaev:2005rs}, this will be explicitly shown within our case in the next section. \newline
On the other hand, the breaking of supersymmetry takes place on the boundaries, whose role is to determine the subspace of possible configurations the fields can lie on, and thus it is a spontaneous breaking, which will be 
checked in section~\ref{susybreaking}.
%
%
%
%
\section{Superfield description}
%
%
%
%
%
 The real formalism is suitable to make contact between an interval approach with boundary mass matrices and an orbifold model with odd bulk masses. 
 However it is convenient to translate this formalism into superfield language where the coupling terms are easily implemented.
 
 Now to recast the action in superfields we will first consider the case $T_0 = T_\pi \equiv T$, while $N_f = \vec p\cdot\vec s_f\,M$, according to what we saw previously and for simplicity we will 
 take\footnote{Notice that we can always do so by means of global rotations of $SU(2)_R$ and $SU(2)_H$, respectively, although $T$ can not be connected with - $T$ by any unitary transformation.} $T = -\sig_3$ and $\vec p_0= (0,0,1)$, the reason for that choice 
 of signs will become clear in a moment.\newline
 Furthermore, the reality constraints (\ref{dirac}), (\ref{scalars}) and (\ref{susyreal}) can be solved as
 \begin{align}\label{indfields}
          \Phi=&  \left(\begin{array}{c}\Phi^1_1 \\ \Phi^1_2 \\ - \Phi^{1*}_2 \\ \Phi^{1*}_1\end{array}\right)\,,\nonumber \\
     \Psi^1 = & \left(\begin{array}{c} \psi^1_L \\ \bar\psi^1_R\end{array}\right)\,,\\
     \Psi^2 = & \left(\begin{array}{c} -\psi^1_R \\ \bar\psi^1_L\end{array}\right)\,, \nonumber
   \end{align}
   \begin{align}
       \chi_1  = & \left(\begin{array}{c} \xi \\ \bar\eta \end{array}\right)\,, \nonumber \\ 
     \chi_2  = & \left(\begin{array}{c} -\eta \\ \bar\xi \end{array}\right)\,,\label{indsusy}
  \end{align}
and, as it is shown in appendix~\ref{A}, the fields can be split into two chiral multiplets according to
 \begin{eqnarray}
      W_c & = & \phi_c + \sqrt2 \theta \psi_c + F_c \theta^2\,,   \\
  W & = & \phi + \sqrt2 \theta \psi + F \theta^2  \,,
       \end{eqnarray}
upon the redefinitions 
\begin{eqnarray}\label{assignations}
  \left(\begin{array}{c} \phi_c \\ \phi\end{array}\right) & \equiv & \left(\begin{array}{c}  -\I \Phi^1_1\\-\I \Phi^{2}_1\end{array}\right)\equiv \varphi \,,\nonumber \\
   \left(\begin{array}{c} F_c \\ F\end{array}\right) & \equiv & \left(\begin{array}{c} - 2F^1_2 + \partial_5\phi^* \\- 2F^{1*}_1 - \partial_5\phi_c^* \end{array}\right)\equiv \mathcal F \,,\nonumber\\
  \left(\begin{array}{c} \psi_c \\ \psi\end{array}\right) & \equiv & \left(\begin{array}{c} \psi^1_L \\ - \psi^1_R\end{array}\right) \,.\nonumber
\end{eqnarray}
 Accordingly, the equations of motion for the auxiliary fields can be compactly expressed as
 \be
 \mathcal F = \e\lsb\varphi'^* + M\sig_3\varphi^*\rsb = \e\lsb\varphi' + M\vec p_0\cdot\vec\sig\,\varphi\rsb^*\,,\label{aux}
 \ee
 with $\e$ the total antisymmetric 2-tensor,
  while the fermionic boundary conditions (\ref{fer}) translate into
 \be
 \lb\id - S\rb\lb\ba{c}\psi_c\\\psi\ea\rb=0\,.
 \ee
 The bosonic sector takes the form
 \begin{align}
& \lb\id - \sig_3\otimes S\rb\lb\ba{c}\varphi\\-\e\varphi^*\ea\rb=0\,,\\
 &\lb\id + \sig_3\otimes S\rb\lsb\lb\ba{c}\varphi'\\-\e\varphi'^*\ea\rb+ M\vec s\cdot\vec p_0 \lb\ba{c}\varphi\\-\e\varphi^*\ea\rb\rsb =0\,,
 \end{align}
 or equivalently\footnote{Now it is clear why we chose $T = -\sig_3$ although such election is not arbitrary since it affects the projector of the supersymmetry parameters (see appendix~\ref{A}).}
 \begin{align}
& \lb\id - S\rb \varphi=0\,,\label{var}\\
 &\lb\id + S\rb\lsb\varphi' + M\vec s\cdot\vec p_0\,\varphi\rsb =0\,.\label{prima}
 \end{align} 
 Finally, using the identity
$$M\,\vec s\cdot\vec p_0\,\id = \frac{1}{2}\lcb S,\mathcal M_0\rcb\,,$$ 
with $\mathcal M_0 = M\vec p_0\cdot\vec\sig$, and that 
$$\varphi=\frac{1}{2}\lb\id + S\rb\varphi\,,$$
Eq.~(\ref{prima}) becomes
$$\lb\id + S\rb\varphi' + \lb S\mathcal M_0 + \mathcal M_0 S\rb\varphi =0\,.$$
 Now adding and subtracting $\mathcal M_0\varphi$ and using Eq.~(\ref{var}) we are left with 
 \be\lb\id+ S\rb\lsb\varphi'+\mathcal M_0\varphi\rsb=0\,.\label{almostF}\ee
 Taking then the complex conjugate of (\ref{almostF}) and using the identities $\e\,S\,\e = S^*$ and $\e^2=-\id$ we finally find
 $$0=\e\lb-\id + S\rb\e\lsb\varphi'+\mathcal M_0\varphi\rsb^*=\e\lb-\id+S\rb\mathcal F\,,$$
and thus
 \be
 \lb\id - S\rb \lb\ba{c}\phi_c\\\phi\ea\rb=\lb\id - S\rb \lb\ba{c}\psi_c\\\psi\ea\rb=\lb\id - S\rb\lb\ba{c}F_c\\ F\ea\rb=0\,,
 \ee
 which explicitly reflects the supersymmetry of the boundary conditions.
 
 Concerning the action one can easily rewrite it as\footnote{For simplicity in the notation we omitted the subscript $f$.}
\begin{align}\label{n1}
\mathcal S = &\int_\Sig\left[\I\bar\psi_c\bar\sig^\m\partial_\m\psi_c + \I\psi\sig^\m\partial_\m\bar\psi - \phi_c^*\Box\phi_c - \phi^*\Box\phi + \abs{F_c}^2 + \abs{F}^2\right.\nonumber\\
&\;\;\;\;\left.+\,F_c\left(-\partial_5 + M\right)\phi + \phi_c\left(-\partial_5 + M\right) F + \psi_c\left(\partial_5 - M\right)\psi+ {\rm h.c.}\right] \nonumber\\
\nonumber\\
&+\int_{\partial\Sig}\left[K - \left(\frac{1}{2}s_3\psi_c\psi - \frac{1}{4}s_+\psi_c\psi_c+\frac{1}{4}s_-\psi\,\psi + {\rm h.c.}\right)\right.\nonumber\\
&\qquad-\left.\frac{1}{2} M \vec p_0\cdot \vec s \,\varphi^\dagger ({\bf 1}  +S) \varphi - \frac{1}{2} (\varphi^\dagger S \varphi)' \right]
\end{align}
where $K$ comes from partial integration and is given by
$$K = \frac{1}{2}\partial_5\left(\abs{\phi_c}^2 + \abs{\phi}^2\right) + M \left(\abs{\phi_c}^2 - \abs{\phi}^2\right) - \frac{1}{2}\left(\psi_c\psi + \bar\psi_c\bar\psi\right) + \phi_c F + \phi_c^* F^*\,,$$
and in addition we have defined
$s_\pm = s_1 \pm \I s_2$ and $\vec s =(s_1,s_2,s_3)$.

Notice that the bulk term of (\ref{n1}) is already $N=1$ invariant without any boundary contribution, which implies that $\mathcal S'_{bd} = \mathcal S_{bd} + \int_{\partial\Sig}\,K$ has to be so. Let us now explicitly check this point.
The fermionic component of $\mathcal S'_{bd}$ is given by
\be
\int_{\partial\Sig}\left[-\frac{1}{2}(1 + s_3)\psi_c\psi + \frac{1}{4}s_+\psi_c\psi_c-\frac{1}{4}s_-\psi\,\psi + {\rm h.c.}\right]
\ee
 while for the bosonic sector we find
\begin{align}\label{bosbd}
   &\int_{\partial\Sig}\left[-\frac{1}{2} M \vec p_0\cdot \vec s \,\varphi^\dagger ({\bf 1}  +S) \varphi + \frac{1}{2} \left(\varphi^\dagger \left({\bf 1}-S\right) \varphi\right)' + M \varphi^\dagger\sig_3\varphi + \phi_c F + \phi_c^* F^* \right]\nonumber\\
   &=\int_{\partial\Sig}\left\{-\frac{1}{2} \varphi^\dagger ({\bf 1}  +S)\left[\varphi'+M \vec p_0\cdot \vec s \,\varphi\right] + \frac{1}{2} \varphi'^\dagger \left({\bf 1}-S\right) \varphi \right.\nonumber\\
   &\qquad\;\;\left.+\,\varphi^\dagger\varphi'+ M \varphi^\dagger\sig_3\varphi + \phi_c F + \phi_c^* F^* \right\}\,.
\end{align}
Using now the boundary conditions 
and the equations of motion for the auxiliary fields 
(\ref{bosbd}) reduces to
\be
\int_{\partial\Sig} \phi^* F_c^* + \phi_c F\ . 
\ee
Then it can be easily checked that 
\begin{align}
\phi^* F_c^* + \phi_c F &= \frac{1}{2}(1+s_3)\left(F_c\phi + \phi_c F\right) + \frac{1}{2}s_-\phi\,F -\frac{1}{2}s_+\phi_c F_c + {\rm h.c.}\nonumber\\
&+\,\frac{1}{2}\varphi^T \e\left({\bf 1} + S\right){\cal F} +\frac{1}{2}{\cal F}^\dagger \e\left({\bf 1} + S^*\right)\varphi^* \,,
\end{align}
 where using the identities $\e\,S\,\e = S^*=S^T$ one immediately realizes that the last two terms separately vanish upon the use of boundary conditions. 
Thus, as claimed, we can write the whole action in terms of superfields as
\begin{align}
           S & =  \int_{\Sig}\di^4\theta \left[\bar W W + \bar W_c W_c\right] - \int_{\Sig}\di^2\theta W_c (\partial_5 - M) W + {\rm h.c.}\nonumber\\  
%
           &+  \int_{\partial\Sig}\di^2\theta\left[\frac{1+s_3}{2}W W_c+ \frac{s_- }{4}W W - \frac{s_+}{4} W_c W_c \right]  + {\rm h.c.}\label{superfield}
   \end{align}
   A comment on gauge charges is in order here. The gauge charges assignment is not the usual one in the sense that 
  if $W$ transforms in the $\mathfrak R$ representation, $W_c$ lies in $\bar{\mathfrak R}$. Instead, one must 
   find a representation of the gauge group where the reality constraints are preserved, as it is done in Ref.~\cite{Diego:2005mu}.

   Recalling that we have taken $\vec p_0 = (0,0,1)$, in order to have a general mass configuration we simply undo the $SU(2)_H$ rotation. Explicitly, (\ref{superfield}) can be rewritten in a compact way as 

\begin{align}
           S & =  \int_{\Sig}\di^4\theta\, \bar{\cal W}{\cal W} - \frac{1}{2}\int_{\Sig}\di^2\theta\left[ {\cal W}^T\e{\cal W}' + M{\cal W}^T\e\,\vec p_0\cdot\vec\sig\,{\cal W} \right]+ {\rm h.c.}\nonumber\\  
%
           &-  \frac{1}{4}\int_{\partial\Sig}\di^2\theta\,{\cal W}^T\e S{\cal W}  + {\rm h.c.}
   \end{align}
   where ${\cal W} = (W_c,\, W)^T$ (already $SU(2)_H$ covariant). Therefore an arbitrary $SU(2)_H$ rotation leaves the kinetic term invariant while the mass term is brought into the generic form
   \be
   M\vec p\cdot\vec\sig\,.\label{generalmatrix}
   \ee
   In fact this is not only the most general mass term compatible with the 4D $N=2$ structure~\cite{Diego:2008zu}, but the most general one compatible with the 5D Lorentz invariance. In terms of a 4-component 5D Dirac spinor the most general mass term can be written as
   \be
    \a \bar\Psi\,\Psi +\b \,\Psi^T C\,\Psi + \b^* \Psi^\dagger C\, \Psi^*\,,\label{genmass}
   \ee
   with $\a\in\mathbb R$ and $C$ the 5D charge conjugation matrix. One can easily check that (\ref{genmass}) expressed in terms of 2-component Weyl spinors precisely yields a mass matrix of the form (\ref{generalmatrix}). 
   \subsection{General boundary term}

In this section we will briefly check that the boundary term previously displayed is indeed on-shell equivalent to the most general boundary term that can be written, which is

%
    \begin{equation}
           \tilde {\cal S}_{\rm bd}  =  \int_{\partial\Sig} d^2 \theta \, [ \frac{\mu}{2} W\, W + \frac{\lambda}{2} W_c W_c + \nu W \,W_c ] + h.c.  \label{bdlagrangian}
   \end{equation}
where $\mu ,\, \lambda$ and $\nu$ are arbitrary complex numbers. The variation of ${\cal S}_{\rm bk} + \tilde {\cal S}_{\rm bd}$ yields the boundary term
\be
\int_{\partial\Sig}\di^2\theta\,\left[\d W_c\left(\lambda W_c + \nu W\right) + \d W\left(\m W + \nu W_c-W_c\right)\right] + {\rm h.c.} \nonumber
\ee
which provides the boundary conditions
\begin{eqnarray}\label{superfieldbound}
         \mu W + \nu W_c - W_c & = & 0 \label{BC1} \\
         \lambda W_c + \nu W & = & 0 \ .\label{BC2}
    \end{eqnarray} 

  One can easily check that in order to not overdetermine the system the complex parameters have to satisfy the relation  
   \begin{equation}\label{overdetermine}
           \mu \ \lambda - (\nu -1) \nu = 0\,,
  \end{equation}
 and that (\ref{BC1})-(\ref{BC2}) are invariant under the redefinitions 
 \be
   W_c \leftrightarrow W,\,\l\leftrightarrow\m,\,\n\leftrightarrow 1-\n\,.\label{redefinicion}
 \ee
 In the special case $\n = 0$ the boundary conditions reduce to 
 \be
 \lcb\ba{l}\l = 0,\,\, \m W - W_c =0 \\
                  \text{\rm or}\\
                  \m =0,\,\, W_c = 0
                  \ea\right.
 \ee
 while the case $\n = 1$ is obtained from the previous one by means of the redefinitions (\ref{redefinicion}). In the general case $\n \notin \lcb 0, 1\rcb$, (\ref{BC1})-(\ref{BC2}) 
  reduce to 
 $$z W_c + W = 0$$
 with $z = \l/\n$. This means that we have a lot of redundancy in the parameters $\n,\m,\l$ since
 only the complex number $z$ plays a role in solving the boundary conditions. Actually by letting $z$ to take any complex value we cover the whole set of boundary conditions including $\n =0$, which corresponds 
 to $z\to \infty$. As a matter of fact, the parameterization 
 \be\n_0 = \frac{1}{2}(1+s_3)\,,\,\m_0 = \frac{1}{2}s_- = \frac{1}{2}\sqrt{1-s_3^2}\,\,\tam{e}^{i\d}\,,\,\l_0 = -\m_0^*\,,\nonumber\ee
 verifies
 $\mu_0 \lambda_0 - (\nu_0 -1) \nu_0 = \frac{1}{4}\left(1-\vec s^2\right) = 0$ and the mapping $z = \sqrt\frac{1-s_3}{1+s_3}\,\tam{e}^{-\I\d}$ covers the whole complex plane.
 %
%
%
%
\section{Supersymmetry breaking by boundary terms}
\label{susybreaking}
 As we saw previously, supersymmetry is broken by the boundary terms  whenever $\vec t_0 \neq \vec t_\pi$ and/or $N_f \neq \vec p\cdot\vec s_f\,M$. The misalignment of the $R$-matrices is equivalent to have  a local 
 $SU(2)_R$ transformation, $e^{\I y\vec\w\cdot\vec \sig}$, such that $T_\pi = e^{\I \pi\vec\w\cdot\vec \sig}\,T_0 \,e^{-\I \pi\vec\w\cdot\vec \sig}$ which is a Scherk-Schwarz like breaking~\cite{Scherk:1978ta,Bagger:2001ep,Diego:2005mu} and therefore a (super) soft breaking. 
 A very elegant proposal consists of 
breaking supersymmetry at the supergravity level via the expectation value 
acquired by some auxiliary field of the supergravity multiplet~\cite{Giudice:1988yz, Marti:2001iw}.

We suggest a very similar breaking mechanism~\cite{DGQ} restricting to the case of flat space $\mathcal M^4 \times I$, where $I$ is the interval $[0,\pi]$, with the metric 
  \be
       ds^2 = \eta_{\mu\nu}dx^\mu dx^\nu - R^2 dy^2\,,
  \ee
  where $R$ is the radion field which parametrizes the compact extra dimension labeled by $y$, which ranges from $0$ to $\pi$. Supersymmetrization of the radion field is given by
  \be
     T = R + \I B_5 + \theta \Psi^5_R + \theta^2 F_T\,, \label{radionsup}
  \ee
  where $B_5$ is the fifth component of the graviphoton, $\Psi^5_R$ is the fifth component of the right-handed gravitino and $F_T$ is a complex auxiliary field. 
The supersymmetric action will be given 
by  
\begin{align}
           S & =  \int_{\Sig}\di^4\theta\, \frac{T + \bar T}{2}\bar{\cal W}{\cal W} - \frac{1}{2}\int_{\Sig}\di^2\theta\left[ {\cal W}^T\e{\cal W}' + M\,T\,{\cal W}^T\e\,\vec p\cdot\vec\sig\,{\cal W} \right]+ {\rm h.c.}\nonumber\\  
%
           &-  \frac{1}{4}\int_{\partial\Sig}\di^2\theta\,{\cal W}^T\e S{\cal W}  + {\rm h.c.}\,.\label{bdlagbreak}
   \end{align}
  Supersymmetry can be spontaneously broken by allowing expectation values for the auxiliary field of the radion
 \be
      \langle T\rangle = R + 2\omega\,\theta^2 \,, \label{T}  
 \ee
 $\omega$ being a dimensionless constant. The bosonic sector of (\ref{bdlagbreak}), disregarding the 4D kinetic term, reads\footnote{The fermions are unaffected by the radion VEV.}
\begin{align}
& \int_\Sig \mathcal F^\dagger\mathcal F -\lcb -\w\,\mathcal F^\dagger\varphi+ \frac{1}{2}\mathcal F^T\e\,\varphi'+\frac{1}{2}\varphi^T\e\,\mathcal F' + M\mathcal F^T\e\,\vec p\cdot\vec\sig\,\varphi\right.\nonumber \\
&\qquad\left.+ \frac{1}{2}M\,\w\,\varphi^T\e\,\vec p\cdot\vec\sig\,\varphi + {\rm h.c.} \rcb -\nonumber\\
&\frac{1}{2}\int_{\pd\Sig} \mathcal F^T\e\,S\varphi+ {\rm h.c.}\,.
\end{align}
Obviously, the boundary conditions are the same as before, that is
\be
\lb\id - S\rb\varphi = \lb\id - S\rb\mathcal F =0\,,\label{bdss}
\ee
and since the new equations of motion for the auxiliary fields are
\be
\mathcal F = \e\lsb\varphi' + M\vec p\cdot\vec\sig\,\varphi\rsb^* - \w\varphi\,,
\ee
it is clear that (\ref{bdss}) is equivalent to the system\footnote{The boundary conditions are the same because we are working in a Hosotani like basis.} 
\begin{align}
& \lb\id - S\rb \varphi=0\,,\\
 &\lb\id + S\rb\lsb\varphi' + M\vec s\cdot\vec p\, \varphi\rsb =0\,,
 \end{align} 
or in the language of real formalism 
 \begin{align}
& \lb\id - \sig_3\otimes S\rb\Phi=0\,,\\
 &\lb\id + \sig_3\otimes S\rb\lsb\Phi'+ M\vec s\cdot\vec p\,\Phi\rsb =0\,.
 \end{align}
In addition, the bulk (bosonic) action can be written as
\be
\int_\Sig -\frac{1}{2}\Phi^\dagger\Box\Phi+\frac{1}{4}\Phi^\dagger D_5^2\Phi + \frac{1}{4}\lb D_5^2\Phi\rb^\dagger \Phi + (\text{\rm mass term}) \,,
\ee
with $D_5 = \pd_5 + \I\w\sig_2$ and thus by a local redefinition
$$\Phi\to \id_H\otimes\lb\tam{e}^{-\I\sig_2 y}\rb_R \Phi\,,$$
the connection is absorbed and the breaking takes place at the boundaries since now $T_\pi = \tam{e}^{\pi\I\sig_2 } T_0 \tam{e}^{-\pi\I\sig_2 }$.
 
  In order to study the nature of the breaking due to the departure of $N_f$ from $M\,\vec s_f\cdot\vec p$ we shall consider the boundary action (\ref{superfield}) plus an effective coupling such that the new boundary term is given by
 \begin{align}
  & -\frac{1}{4}\int_{\partial\Sig}\di^2\theta\mathcal W^T\e\,S\mathcal W  + {\rm h.c.}\nonumber \\
  & - \frac{1}{\L^3}\int_{\partial\Sig}\di^4\theta \frac{1}{2}\lb\bar{\mathcal N'_f} \,\mathcal N_f + \bar{\mathcal N_f}\,\mathcal N'_f\rb\bar{\mathcal W}\,\mathcal W\,, \label{bdbreaking}
 \end{align}
	 with $\L$ the scale of the cutoff and $\mathcal N_f, \mathcal N_f'$ localized superfields whose auxiliary fields acquire VEVs, say $F_f, F_f'$, such that $\frac{{\bar F_f' F_f}}{\L^3} =  N_f$. Now the (bosonic) boundary 
	 conditions\footnote{The fermionic boundary conditions are unaffected.} turn into
 \begin{align}
   &\left(\id - S_f \right) \varphi = 0\,,\label{primera}\\
   &\left(\id - S_f \right)\mathcal F-2 N_f \e\,\varphi^*  = 0\,,
 \end{align}
with $\mathcal F$ given by (\ref{aux}).
Finally, using the identity $\e\,S\,\e = S^*$ we are left with
  \begin{align}
   &\left(\id - S_f \right) \varphi = 0\,,\\
  &\left(\id + S_f \right) \left[\pd_5 + \vec s_f\cdot\vec p\,M + N_f\right] \varphi  = 0\,,
 \end{align}
This shows explicitly that this breaking has a soft nature, a result which the calculation of the radiative corrections coming from such breaking term to the Higgs mass coupling strongly suggests~\cite{Diego:2005mu}. 
%
%
\section{Conclusions}
The main objective of the present work was to explicitly show the translation into superfield language of a model for ElectroWeak Symmetry Breaking (EWSB) developed in component fields subject to reality constraints. 
Albeit the model was proven to be (on-shell) supersymmetric under certain bulk-brane configurations, that being broken by boundary terms, it was not totally evident how to 
develop an equivalent off-shell supersymmetric model. To build such a dictionary was motivated by several reasons:
The model predicted a very interesting scenario where a tachyon mode for the Higgs was present at tree level. This opened a chance for the EWSB to be triggered by the negative top-stop corrections, since the negative 
squared mass could partially cancel the positive gauge corrections. For that, however, an exhaustive study of the quantum behavior of the model is needed but, unfortunately, to embed interacting terms within the real formalism 
is not an easy task. On the other hand, the breaking of supersymmetry comes from the misalignment between several bulk-brane parameters, one of them being easily identified with a Scherk-Schwarz
like breaking, and coming both from boundary terms it indicates a spontaneous mechanism, nevertheless, a explicit translation into superfield formalism helps to clarify its nature.\newline
It is worth remarking, however, that the dictionary we have developed so far does not mean to be neither a formal proof nor a consistent extension of the model, at any level. For that, among other aspects, one should 
justify the presence of the spurion fields breaking the supersymmetry through their vacuum expectation values.
 \section{Acknowledgements} 
The author would like to thank Gero v. Gersdorff and Mariano Quir\'os for their useful comments and for their direct help while making some of the calculations. This acknowledgement is as well extendible to 
Antonio Delgado  for his helpful comments and remarks while redacting the paper.  
 \newpage  
\appendix
\section{Supersymmetry of boundary conditions}\label{B}
 The (on-shell) variations of the boundary conditions, Eqs.~(\ref{fer}-\ref{bos}-\ref{bos'}), are given by
\be
\lb\id + R\rb\d_\chi\Phi =  -\bar\chi\gd\lb\I\gd-S\rb\Psi +\I\lb\bar\chi\I\gd + T\,\bar\chi\rb S\Psi\,,\nonumber
\ee
\begin{align}
\lb-\id+R\rb\lsb\d_\chi\Phi' + N\d_\chi\Phi\rsb =&\I\lb\bar\chi\I\gd+T\,\bar\chi\rb\mathcal M\Psi +T\,\bar\chi\gd\mathcal M\lb\I\gd-S\rb\Psi \nonumber\\
 &-\lb\bar\chi\I\gd + T\,\bar\chi\rb\spd\Psi-\I T\,\bar\chi\gd\spd\lb\I\gd - S\rb\Psi\nonumber\\
&+\I \lb N - M \vec p\cdot\vec s\rb T\,\bar\chi\lb\I\gd + S\rb\Psi\ \nonumber\\
&-\I N\lb\bar\chi\I\gd+T\,\bar\chi\rb\I\gd\Psi\,,  \nonumber
\end{align}
\begin{align}
  \lb-\I\g^{\dot 5}+ S\rb\d_\chi\Psi = & -\g^\m\lb\I\g^{\dot 5}\chi - \chi\,T\rb\pd_\m\Phi - \spd\chi\,T\lb\id + R\rb\Phi \nonumber\\
  &+\lb\I\g^{\dot 5}\chi - \chi\,T\rb\mathcal M\Phi + \chi\,T\mathcal M\lb\id + R\rb\Phi\nonumber\\
  &+ \I\gd\chi\,T\lb-\id+R\rb\lsb\Phi' + M\vec p\cdot\vec s\,\Phi\rsb \nonumber\\
  &+\gd\lb\I\gd\chi-\chi\,T\rb\Phi' \,,\nonumber
\end{align}
where we have used the (bulk) equations of motion for the fermions as well as those for the auxiliary fields. Furthermore, we have omitted the $R$ and $H$ indices as well as the subscript $f$. The above variations cancel upon the restrictions 
\be N_f = M\vec p\cdot\vec s_f\,,\qquad \lb\I\gd - T^T\rb\chi=0\,.\label{susycond}\ee
%

\section{N=1 splitting}\label{A}
To complete the dictionary between the real and the superfield descriptions we will briefly give the splitting of the 5D hypermultiplet into 4D superfield pieces. Being $T = -\sig_3$ the projection on the supersymmetry parameters, Eq.~(\ref{susycond}), reads
\be\lb\id + \sig_3\otimes\I\gd\rb\lb\ba{c}\chi_1\\\chi_2\ea\rb = \lb\ba{c}\lb\id +\g^5\rb\chi_1\\\lb\id-\g^5\rb\chi_2\ea\rb=0\,,\ee
with 
$$\chi_1=\left(\begin{array}{c} \xi \\ \bar\eta \end{array}\right)\,,$$
 as a consequence $\xi = 0$ and the supersymmetric transformations, Eq.~(\ref{susytransf}), can be written as
\begin{eqnarray}\label{etaR}
   \delta_\eta (-i\Phi^1_2)^*  & = &  \eta(- \psi^1_R) \nonumber \\
   \delta_\eta (- \psi^1_R) & = & -i\sigma^\mu\bar\eta\partial_\mu (-i\Phi^1_2)^* + \eta (- 2F^1_1 + i \partial_5\Phi^1_1)^*  \nonumber \\
   \delta_\eta (-2F^1_1 + i\partial_5\Phi^1_1)^* & = & - i \bar\eta\bar\sigma^\mu \partial_\mu (- \psi^1_R)
\end{eqnarray}
\begin{eqnarray}\label{etaL}
   \delta_\eta (-i\Phi^1_1)  & = & \eta( \psi^1_L) \nonumber \\
   \delta_\eta ( \psi^1_L) & = & -i\sigma^\mu\bar\eta\partial_\mu (-i\Phi^1_1) + \eta (- 2F^1_2 - i \partial_5\Phi^1_2)  \nonumber \\
   \delta_\eta (- 2F^1_2 - i\partial_5\Phi^1_2) & = &  - i \bar\eta\bar\sigma^\mu \partial_\mu (-i \psi^1_L)
\end{eqnarray}
which corresponds to the pair of chiral superfields~\cite{wb}
 \begin{eqnarray}
       W & = & \I\Phi_2^{1*} + \sqrt2 \,\theta\lb-\psi_R^1\rb + \lb-2 F_1^{1*} -\I\pd_5\Phi_1^{1*}\rb \theta^2  \,,\\
       W_c & = & \lb-\I\Phi_1^1\rb + \sqrt2 \,\theta\psi^1_L + \lb-2 F_2^1 - \I\pd_5\Phi_2^1\rb \theta^2\,.
 \end{eqnarray}
%

%

\newpage
\end{document}